\documentclass[preprint,showpacs,showkeys,preprintnumbers,amsmath,amssymb]{revtex4}

\usepackage{graphicx}
\usepackage{dcolumn}
\usepackage{bm}

\begin{document}

\title{Asymptotic regime in \emph{N} random
interacting species}

\author{A. Fiasconaro\footnote{e-mail: afiasconaro@gip.dft.unipa.it},
D. Valenti and B. Spagnolo}
 \affiliation{Dipartimento di Fisica e
Tecnologie Relative, Universit\`a di Palermo\\ and INFM-CNR, Group
of Interdisciplinary Physics\footnote {Electronic
address: http://gip.dft.unipa.it},
Viale delle Scienze, I-90128 Palermo, Italy\smallskip}
\date{\today}

\begin{abstract}
The asymptotic regime of a complex ecosystem with \emph{N }random
interacting species and in the presence of an external
multiplicative noise is analyzed. We find the role of the external
noise on the long time probability distribution of the
\emph{$i^{th}$} density species, the extinction of species and the
local field acting on the \emph{$i^{th}$} population. We analyze
in detail the transient dynamics of this field and the cavity
field, which is the field acting on the $i^{th}$ species when this
is absent. We find that the presence or the absence of some
population give different asymptotic distributions of these
fields.
\end{abstract}


\maketitle

\section{Introduction}
In recent years great attention has been devoted to population
dynamics modelled by generalized Lotka-Volterra
systems~\cite{McKane}. Ecosystems are a classic example of complex
systems, which became object of study as well by biologists as by
physicists~\cite{Mur93,Katja04}. Tools developed in the context of
nonequilibrium statistical physics to analyze nonequilibrium
nonlinear physical systems provide new insights and at the same
time new approaches to the comprehension of the properties of
biological and many body systems. A key aspect to understand the
complex behavior of ecosystems is the role of the external noise
on the dynamics of such systems. Noise-induced effects in
population dynamics, such as pattern formation~\cite{Soc01,Fia04},
stochastic resonance, noise delayed extinction, quasi periodic
oscillations etc... have been investigated with increasing
interest~\cite{Spa03,Val04,Sci99,Spa04}. The dynamical behavior of
ecological systems of interacting species evolves towards the
equilibrium states through the slow process of nonlinear
relaxation, which is strongly dependent on the random interaction
between the species, the initial conditions and the random
interaction with environment. Moreover biological evolution
presents the same fundamental ingredient that characterizes
physical systems far from equilibrium in their route to
equilibrium, namely the disorder-order transition. Different
models of evolution are reported in literature, which are useful
to describe a lot of dynamical population
problems~\cite{Mos00,Pel97}. Among them it is worthwhile to cite
two bit-string models of population dynamics, namely the Eigen
model ~\cite{Eig89} and the Penna model ~\cite{Pen95}.

The mathematical model here used to analyze the dynamics of $N$
biological species with spatially homogeneous densities is the
generalized Lotka-Volterra system with a Malthus-Verhulst
modelling of the self regulation mechanism and with the addition
of an external multiplicative noise source~\cite{Ciu96,Spa02}. We
obtain the asymptotic behaviors of the probability distribution of
the populations for different values of external noise intensity.
To analyze the role of the external noise on the transient
dynamics of the species we focus on the long time distribution of
the local field, that is the interaction term in the dynamical
equation of the \emph{$i^{th}$} species that takes the influence
of all other species into account. We find that the presence or
the absence of some population give different asymptotic
distributions of the local field and of the cavity field (field
acting on the \emph{$i^{th}$} species when this is absent) in the
absence of external noise. When the noise is switched on the
asymptotic local and cavity fields tend to overlap and
approximately superimpose each other for very high noise
intensity. Finally the long time evolution of the average number
of the extinct species is reported for different values of the
multiplicative noise intensity.

\section{The model}

The dynamical evolution of our ecosystem composed by \textit{$N$}
interacting species in a noisy environment (climate, disease,
etc...) is described by the following generalized Lotka-Volterra
equations with a multiplicative noise, in the framework of Ito
stochastic calculus

\begin{equation}
d n_i(t) = \left[ \left(g_i(n_i(t)) + \sum_{j\neq i} J_{ij}n_j(t)
\right)dt + \sqrt{\epsilon} dw_i\right] n_i(t)\mbox{,} \enspace
\qquad i = 1,...,N
\label{langevin}
\end{equation}
where \textit{$n_i(t) \geq 0$} is the population density of the
\textit{$i^{th}$} species at time \textit{$t$} and the function
$g_i(n_i(t))$

\begin{equation}
g_i(n_i(t)) = \left(\alpha + \frac{\epsilon}{2} \right) - n_i(t),
\label{g function}
\end{equation}
describes the development of the \textit{$i^{th}$} species without
interacting with other species. In Eq.~(\ref{langevin}),
\textit{$\alpha$} is the growth parameter, the interaction matrix
\textit{$J_{ij}$} models the interaction between different species
($i\neq j$), and \textit{$w_i$} is the Wiener process whose
increment \textit{$dw_i$} satisfy the usual statistical properties
\emph{$ \langle dw_i(t) \rangle \thinspace = \thinspace 0$}, and
\emph{$\langle dw_i(t)dw_j(t^{\prime})\rangle \thinspace =
\thinspace \delta_{ij}\delta(t-t^{\prime}) dt$}. We consider an
asymmetric interaction matrix \emph{$J_{ij}$}, whose elements are
randomly distributed according to a Gaussian distribution with
\emph{$\langle J_{ij}\rangle = 0$}, \emph{$\langle J_{ij}
J_{ji}\rangle = 0$}, and \emph{$\sigma^2_j = J^2/N$}. Therefore
our ecosystem contains 50$\%$ of prey-predator interactions
(\emph{$J_{ij}< 0$} and \emph{$J_{ji} > 0$}), 25$\%$ competitive
interactions (\emph{$J_{ij}>0$} and \emph{$J_{ji}>0$}), and 25$\%$
symbiotic interactions (\emph{$J_{ij}<0$} and \emph{$J_{ji}<0$}).
We consider all species equivalent so that the characteristic
parameters of the ecosystem are independent of the species. The
formal solution of Eq.~(\ref{langevin}) is

\begin{equation} n_i(t) = \frac{n_i(0) z_i (t)} {1+
n_i(0) \int_{0}^{t}dt^{\prime}  z_i (t^{\prime})}\;,
\label{sol-langevin}
\end{equation}
%
%
where
\begin{equation}
  z_i(t) = exp\left[\alpha t +\sqrt{\epsilon} w_i(t) +
\int_{0}^{t} dt^{\prime}h_{i, loc}(t^{\prime})\right]\; .
\label{z-process}
\end{equation}
The term $h_{i,loc}(t) = \sum_{j\neq i}J_{ij}n_j(t)$ is the local
field acting on the \textit{$i^{th}$} population and represents
the influence of other species on the differential growth rate. We
note that the dynamical behavior of the \textit{$i^{th}$}
population depends on the time integral of the process
\textit{$z_i(t)$} and the time integral of the local field.

In the absence of external noise ($\epsilon = 0$), for a large
number of interacting species we can assume that the local field
\textit{$h_i(t)$} is Gaussian with zero mean and variance

\begin{equation}
\sigma_{h_{i,loc}}^2 = \Sigma_{j,k} \langle J_{ij} J_{ik} \rangle
\langle n_j n_k \rangle = J^2 \langle n_i^2 \rangle \;,
\;\;\mathrm{ with }\;\; \langle J_{ij} J_{ik} \rangle =
\delta_{jk}\frac{J^2}{N}\;. \label{sigma h}
\end{equation}
As a consequence, in the absence of external
 noise, from the fixed-point equation $n_i(\alpha -
n_i + h_i) = 0$, the stationary probability distribution of the
populations is the sum of a truncated Gaussian distribution at
$n_i =0$ ($n_i > 0$ always) and a delta function for extinct
species. The initial values of the populations $n_i(0)$ have also
Gaussian distribution with mean value $\langle n_i(0)\rangle = 1$,
and variance $\sigma^2_{n(0)} = 0.03$. The interaction strength
$J$ determines two different dynamical behaviors of the ecosystem.
Above a critical value \emph{$J_c = 1.1$}, the system is unstable
and at least one of the populations diverges. Below $J_c$ the
system is stable and asymptotically reaches an equilibrium state.
The equilibrium values of the populations depend both on their
initial values and on the interaction matrix. If we consider a
quenched random interaction matrix, the ecosystem has a great
number of equilibrium configurations, each one with its attraction
basin. For an interaction strength $J = 1$ and an intrinsic growth
parameter $\alpha = 1$ we obtain: $\langle n_i \rangle = 1.4387,
\langle n^{2}_i \rangle = 4.514,$ and $\sigma^{2}_{n_i} = 2.44$.
These values agree with that obtained from numerical simulation of
Eq.~(\ref{langevin}).

\begin{figure}[htb]
 \includegraphics[width=11 cm]{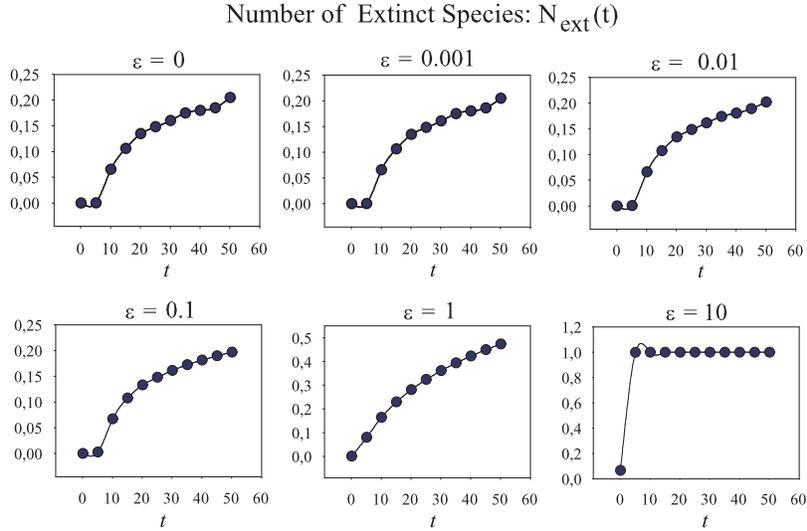}
\caption{Time evolution of the average number of extinct species
for different noise intensities.}
 \label{extinct}
\end{figure}

In the presence of external noise ($\epsilon \neq 0$) we calculate
long time probability distribution for different values of the
noise intensity. These are shown in the following
Fig.~\ref{density}.
\begin{figure}[htb]
 \includegraphics[width=14 cm]{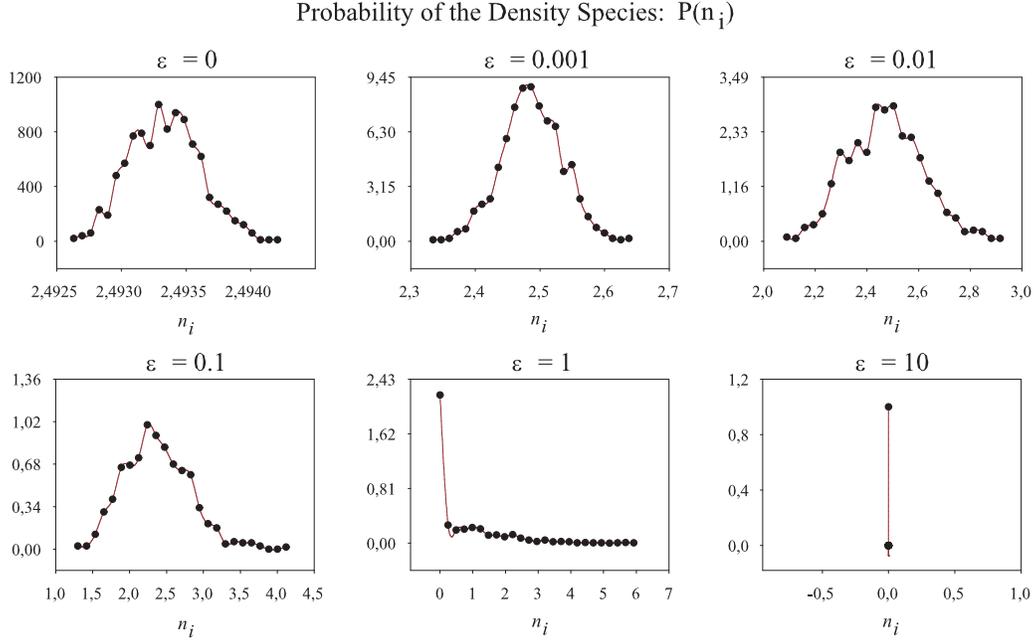}
\caption{Probability distribution for the species densities. The
values of the external noise intensity are $\epsilon = 0, 0.001,
0.01, 0.1, 1, 10$.}
 \label{density}
\end{figure}
For increasing external noise intensity we obtain a larger
probability distribution with a lower maximum (see the different
scales in Figs.~\ref{density} for different noise intensity
values). The distribution becomes asymmetric for $\epsilon = 0.1$
and tends to become a truncated delta function around the zero
value ($P(n_i) = \delta(n_i)$ for $n_i > 0$, and $P(n_i) = 0$ for
$n_i \leq 0$), for further increasing noise intensity. The role of
the multiplicative noise is to act as an absorbing barrier at $n_i
= 0$~\cite{Ciu96}. To confirm this picture we calculate the time
evolution of the average number of extinct species for different
noise intensities. This time behavior is shown in
Fig.~\ref{extinct}. We see that this number increases with noise
intensity, and after the value $\epsilon = 0.1$ reaches quickly
the normalized maximum value at $\epsilon = 10$.

To analyze in more detail the influence of each species on the
long time dynamics of the ecosystem we calculate in this regime
the \textit{local} field $h_{i,loc}(t)$ and the \textit{cavity}
field $h_{i,cav}(t)$, which is the field acting on the {\em
$i^{th}$} population when this population is absent. The
probability distributions for both local and cavity fields have
been obtained by simulations for different species in the presence
and in absence of external noise. The results are shown in the
next section (see Fig.~\ref{fields}). We found that the
probability distributions of the cavity fields differ
substantially from those of local fields for the same species,
while in the presence of noise the two fields overlap. To quantify
this overlap between the probabilities distributions of the two
fields we define an \emph{overlap coefficient} $\lambda(t)$, which
is the distance between the average values of the two
distributions, normalized to their widths

\begin{equation}
 \lambda(t) = \frac{\bar{h}_{i,loc}-\bar{h}_{i,cav}}
 {\sqrt{\sigma_{i,loc}^2+\sigma_{i,cav}^2}} =
 \frac{d_h(t)}{\sigma_d(t)}\;,
 \label{overlap_eq}
\end{equation}
where
\begin{equation}
d_h(t) = \bar{h}_{i,loc}-\bar{h}_{i,cav}\;,\;\;\;\sigma_d^2(t) =
\sigma_{i,loc}^2(t) + \sigma_{i,cav}^2(t) \;.
\end{equation}
With this definition the distributions start to overlap
significantly for $|\lambda| \lesssim 1$, and become strongly
overlapping for $|\lambda| \ll 1$.

\section{Results and Comments}

In the calculation the following parameters have been used:
$\alpha=1.2$, $J=1$, $\sigma_J^2= 0.005$, $N = 200$; the number of
averaging experiment used is $N_{exp}=1000$. Concerning the
initial condition the parameters are: $\langle n_i \rangle = 1$,
$\sigma_{n_o}^2 = 0.03$. The dynamics of various species are
different even if they are equivalent according to the parameters
in the dynamical Eq.~(\ref{langevin}). However we note that to
change the species index by fixing the random matrix or to change
the random matrix by fixing the species index is equivalent as
regards the asymptotic dynamical regime.
\begin{figure}[htb]
\includegraphics[width=14 cm]{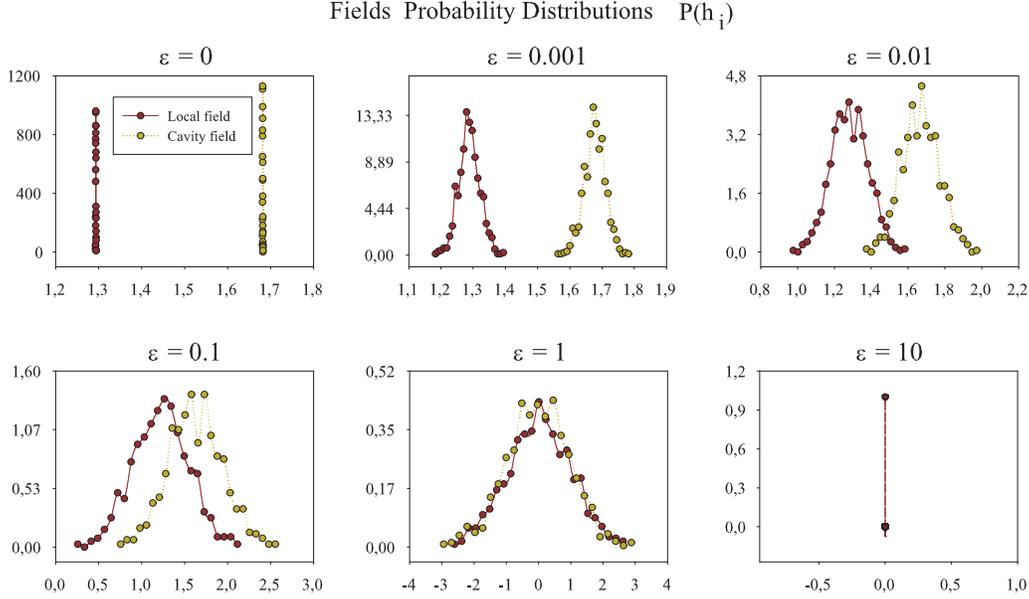}
\caption{Probability distribution of both the local (black
circles)vand the cavity (white circles) fields for various values
of noise intensity $\epsilon = 0, 0.001, 0.01, 0.1, 1$. The graph
are taken at the time $t=50$ unit steps.}
 \hfill
\label{fields}
\end{figure}
Fig.~\ref{fields} shows for various noise intensities the local
and cavity fields probability distributions at time $t = 50$ (a.
u.). For noiseless dynamics the distributions of the fields for
the species $i = 3$ appear very narrow around their mean values
and very spaced each other. By increasing the noise intensity, we
observe an equal enlargement of the two distributions, maintaining
however the same mean values. At $\epsilon=0.1$ the two
distributions start to overlap until, for stronger noise intensity
($\epsilon=1$), they superimpose each other. The \emph{overlap
coefficient} $\lambda(t)$ is equal to zero. The noise makes
equivalent all the species in the asymptotic regime and the
absence of some species doesn't contribute to any changes in the
dynamics of all other species. The last plot in Fig.~\ref{fields}
($\epsilon=10$) gives a delta distribution around zero. This means
that, at the time considered in our simulation ($t=50$) and for
this noise intensity, all the species are extinct (see also
Fig.~\ref{extinct}).

\begin{figure}[htb]
 \includegraphics[width=11 cm]{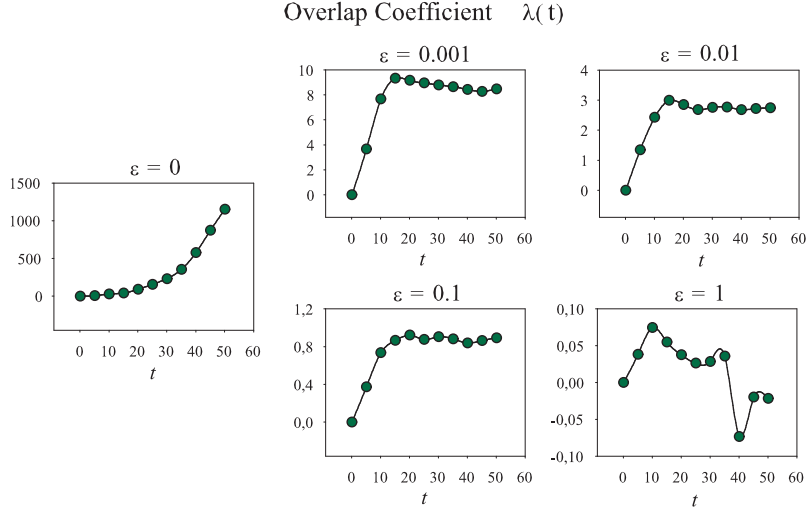}
\caption{Time evolution of the overlap coefficient $\lambda(t)$
between the distributions of local and cavity fields for different
noise intensities. The increasing behavior of the overlap
coefficient as a function of time, in the noiseless case, is due
to the narrowing of the two fields distributions ($\sigma_{loc},
\sigma_{cav} \rightarrow 0$) towards a $\delta$-function. For a
noisy environment the distributions tend to enlarge, decreasing
the value of the overlap coefficient $\lambda(t)$.}
\label{overlap}
\end{figure}

\begin{figure}[htb]
 \includegraphics[width=11 cm]{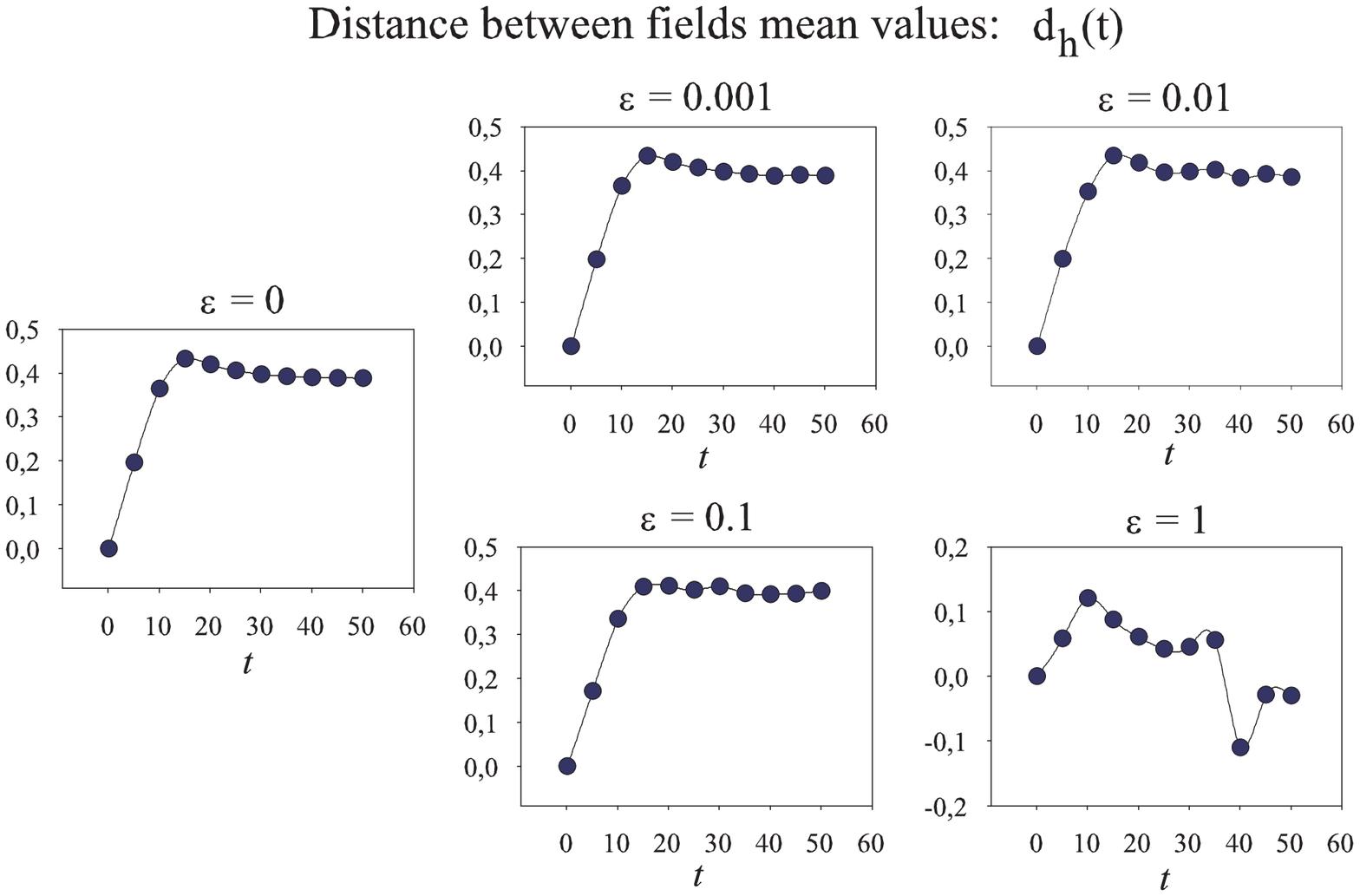}
 \caption{Time evolution of the distance $d_h(t)$ between the mean values of
the fields distributions for different noise intensities, namely
$\epsilon$ = $0, 0.001, 0.01, 0.1,1$.}
 \label{distance}
\end{figure}

\begin{figure}[htb]
 \includegraphics[width=11 cm]{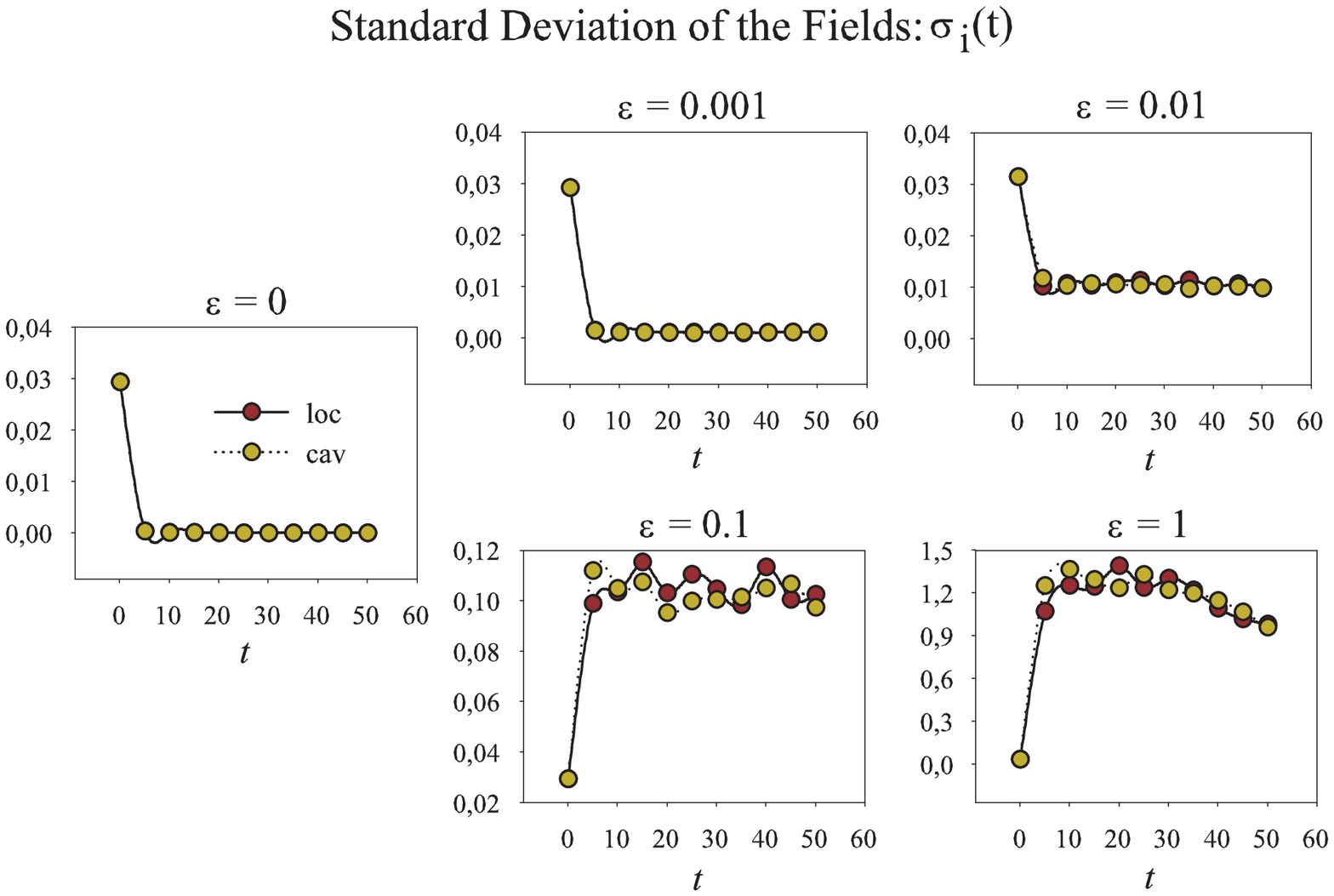}
 \caption{Time evolution of the standard deviation
 $\sigma_{i,loc}^2$ (grey circles)and $\sigma_{i,cav}^2$ (black circles) of the field
distributions for the same noise intensities of
Fig.~\ref{distance}.}
 \label{sigma}
\end{figure}

The detailed time evolution of the overlap of the two
distributions can be seen from Fig.~\ref{overlap}, where it is
plotted the coefficient defined in Eq.~(\ref{overlap_eq}). For
$\epsilon =0$ the $\lambda(t)$ coefficient increases with time.
This is due to the different time behavior of the distance between
the mean values of the field distributions and of their standard
deviation $\sigma_{i,loc}^2$ and $\sigma_{i,cav}^2$. The distance
$d_h(t)$ = $\bar{h}_{i,loc}-\bar{h}_{i,cav}$ is almost constant in
time, except a rapid initial transient (see Fig.~\ref{distance},
$\epsilon =0$), but the corresponding evolution of the
distribution widths decreases rapidly in time. This effect is due
to the quenched random matrix. This behavior remains unchanged
until the noise intensity reaches the value of $\epsilon = 0.01$.
>From this value of external noise intensity some differences start
to be visible (see Figs.~\ref{overlap},~\ref{distance}
and~\ref{sigma}), and at $\epsilon = 1$, after some fluctuations,
both the distance $d_h(t)$ and the overlap coefficient
$\lambda(t)$ reach a value close to zero. The two field
distributions are totally overlapped.

It is worthwhile to note that the behaviors shown in
Figs.~\ref{fields},~\ref{overlap},~\ref{distance} and~\ref{sigma}
have been found for some species, and changing the species
different evolutions of the distribution dynamics appear and in
particular of the $\lambda(t)$ coefficient. This is due to
complexity of our ecosystem and to the extinction process during
the transient dynamics. Moreover this strange behavior, found for
some populations and in the asymptotic regime, is reminiscent of
the phase transition phenomenon~\cite{Ciu88}, and it is related to
the following peculiarities of our dynamical system: (i) all the
populations are positive; (ii) different initial conditions drive
the ecosystem into different attraction basins; and (iii) the
complex structure of the attraction basins. While in the presence
of noise all the populations seem to be equivalent in the long
time regime, some populations, in the absence of external noise,
have an asymptotical dynamical behavior such that they
significantly influence the dynamics of other species. A more
detailed analysis on these points will be done in a forthcoming
paper.

\section{Conclusions}

We analyzed the asymptotic regime of an ecosystem composed by
\emph{N} interacting species in the presence of multiplicative
noise. We find the role of the noise on the asymptotic probability
distribution of populations and on the extinction process.
Concerning the local and the cavity fields, a phase transition
like phenomenon is observed. Their probability distributions tend
to overlap each other in the presence of external noise, reaching
strong overlap for high noise intensity ($|\lambda(t)| \approx
0$), while they are separated ($|\lambda(t)| > 1$) in the absence
of noise. This phenomenon can be ascribed to the complexity of our
ecosystem.

\section*{Acknowledgements}
This work was supported by MIUR and INFM-CNR.

\end{document}